\documentclass[letterpaper,10pt]{article} 

\usepackage{osameet3} 

\usepackage{amsmath,amssymb}
\usepackage{tikz,tkz-euclide}
\usetikzlibrary{matrix,positioning}
\usepackage{adjustbox}
\usepackage{graphicx}
\usepackage{tikz}
\usepackage{color}
\usepackage{lipsum,adjustbox}
\usetikzlibrary{decorations}
\usepackage{pgfplots,siunitx}
\usepackage{pgfplotstable}
\usetikzlibrary{decorations.pathreplacing}
\usetikzlibrary{shapes,arrows,shadows}
\usetikzlibrary{calc,positioning}
\usepackage{tikz,tkz-euclide}
\usepackage{balance}
\usepackage{cite}
\usepackage{multicol}
\usepackage[colorlinks=true,bookmarks=false,citecolor=blue,urlcolor=blue]{hyperref} 
\begin{document}

\title{Multi-dimensional Short Blocklength Probabilistic Shaping for Digital Subcarrier Multiplexing Systems}

\author{Abdelkerim Amari and André Richter}
\address{VPIphotonics GmbH, Carnotstr. 6, 10587, Berlin, Germany}
\email{abdelkerim.amari@vpiphotonics.com}

\copyrightyear{2021}

\begin{abstract}
We propose multi-dimensional short blocklength probabilistic shaping to increase the nonlinear tolerance gain in digital subcarrier multiplexing transmission systems and demonstrate an improvement in performance compared to lower dimensional formats.
\end{abstract}

\section{Introduction}
Probabilistic shaping has been considerably investigated and employed in optical communication systems for rate adaptation and performance improvement \cite{ps2,me1,pss3}. Short blocklength shaping has attracted special attention due to its nonlinear tolerance gain with respect to the long blocklength case and uniform signaling \cite{me1,me3,ps3}, and it is more interesting in case of low rate loss shaping schemes like sphere shaping \cite{me1,ps4}. This makes short blocklength shaping a promising approach to be considered for practical hardware implementation because of the low latency and complexity in comparison with long blocklength shaping and nonlinearity compensation via digital signal processing \cite{me2}.
The nonlinear tolerance gain has been shown to be dependent on the temporal structure of the shaped signal \cite{psnl1}. Intra-distribution matching (DM) pairing, in which the in-phase ($I$) and quadrature ($Q$) components of the Quadrature amplitude modulation (QAM) symbol are formed from two consecutive amplitudes of the same shaping block, provides better performance than inter-DM pairing, i.e. $I$ and $Q$ are taken from different shaping blocks \cite{psnl1}. The idea of intra-DM mapping has been extended to form a shell mapping-based 4-dimension (4D) super symbol by considering the polarization dimension, i.e. the $I$ and $Q$ components of the two polarizations at the same time slots are formed using the same shaping block \cite{psnl2}. 
To enable high bit rate for long-haul optical systems, digital subcarrier multiplexing (DSCM) has been employed \cite{dsm1}. 
The availability of the adjacent subcarriers' information in DSCM allows joint optimization of the signals on the digital subcarriers. DSCM has been recently considered along with probabilistic shaping in an $800$G long-haul transmission system \cite{dsm3}. 

In this work, we propose a multi-dimensional probabilistic shaping for performance improvement in high-speed long-haul systems. We consider intra-DM mapping on polarization and frequency dimensions in the context of DSCM system. A performance improvement in terms of nonlinear tolerance and achievable information rate (AIR) is observed with respect to the conventional $2$D intra-DM pairing and the $4$D super symbol intra-DM mapping. 
\section{Multi-Dimensional Intra-DM Shaping Principle}
Multi-dimensional probabilistic shaping is proposed based on the polarization and frequency dimensions to increase the nonlinear tolerance gain for short blocklength shaping schemes. The shaping is implemented via enumerative sphere shaping (ESS) due to its low rate loss at short blocklength in comparison with constant composition DM (CCDM) \cite{me1}.
The ESS-based shaper is not a distribution matcher as is the case for the CCDM \cite{ps5}. However, we use the term 'DM' in this work because it is commonly used in short blocklength shaping references.

The principle of the proposed multi-dimensional intra-DM shaping is depicted in Fig. \ref{fig:2}(a). Firstly, an ESS block with a shaping rate $R_s = k/n$ converts each block of $k$ information bits into $n$ shaped amplitudes. According to the number of information bits, $m$ shaping blocks with size $n$ $(B_1,B_2,\ldots,B_m)$ are generated at the output of the ESS block. 
Then, the $I$ and $Q$ components of each QAM symbol for both polarizations of all subcarriers are formed using consecutive amplitudes of the same shaping block length. For example, as it is shown on the bottom of Fig. \ref{fig:2}, the shaping block $B_1$ is divided into $16$ subblocks, where $X_I$ and $X_Q$ of subcarrier $1$ are two consecutive amplitudes of $B_1$, and similarly for $Y_I$ and $Y_Q$ and all the $I$ and $Q$ components of the $4$ subcarriers. The proposed multi-dimensional intra-DM shaping is compared with the conventional $2$D intra-DM shaping and the $4$D super symbol intra-DM shaping \cite{psnl2}. The shaping rate considered is $R_s=1.73$ [bits/amp.] and the blocklength is $n=360$, which provides a good trade-off between linear shaping gain and nonlinear tolerance (see Fig. $7$ in \cite{me1}). 

\section{Simulation Setup and Results}
\vspace{-0.1cm}
\begin{figure*}[tbp]
  \begin{minipage}[t]{0.49\textwidth}
	\centering	
	\includegraphics[width=1\linewidth]{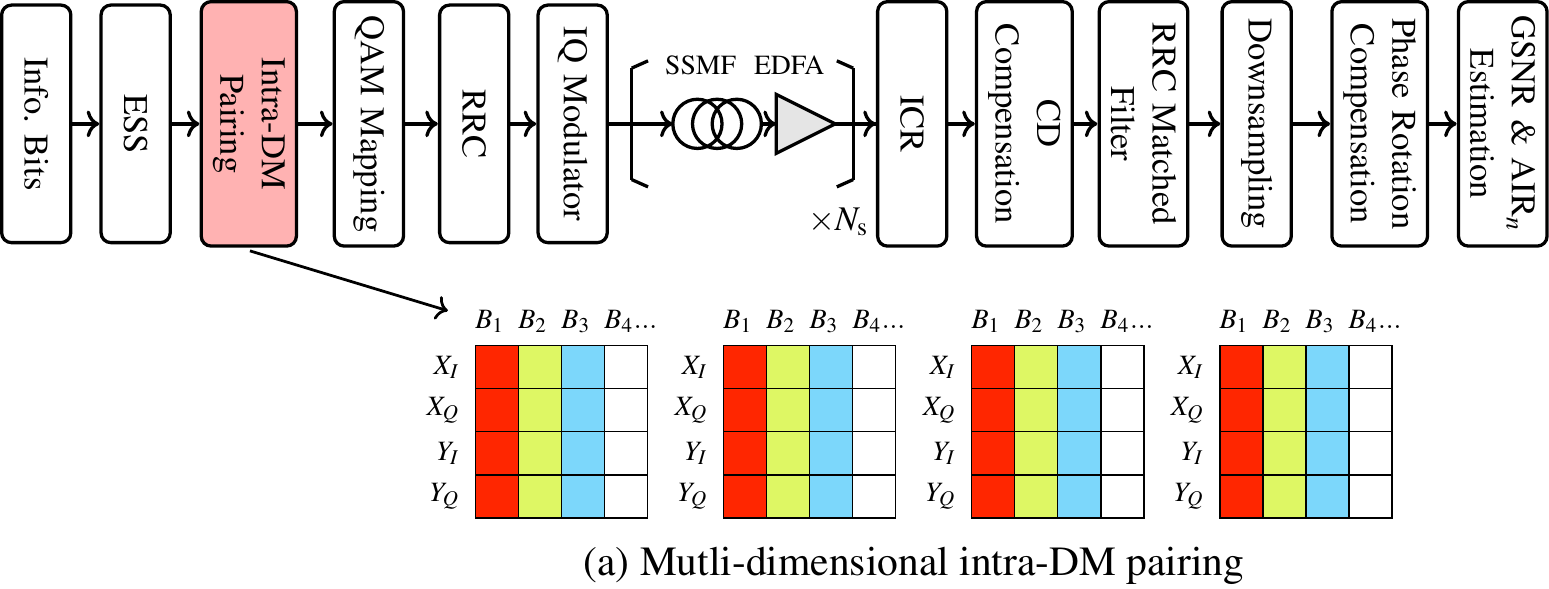}
    \vspace{-0.85cm}
		\caption{Transmission system: (a) Coding principle, (b) GSNR vs input power, (c) AIR vs input power.	}
		\label{fig:2}
  \end{minipage}
  \begin{minipage}[t]{0.25\textwidth}
    \includegraphics[width=1\linewidth]{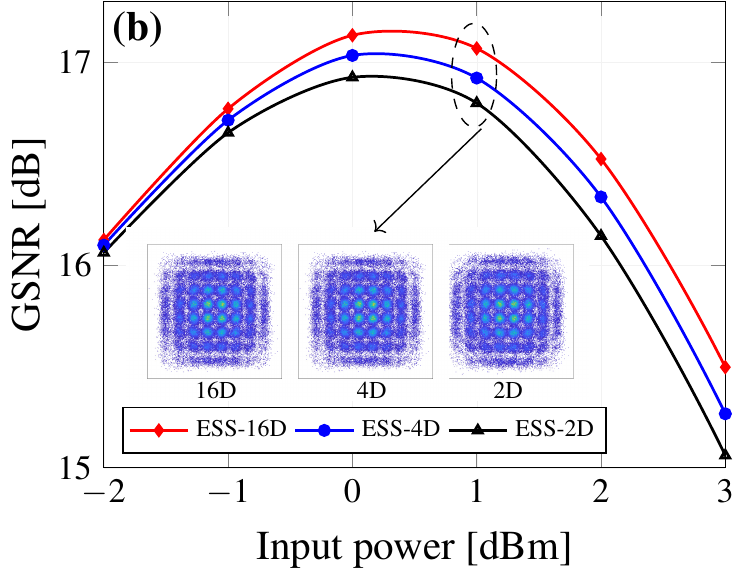}%
    \end{minipage}
    \begin{minipage}[t]{0.25\textwidth}
    \includegraphics[width=1.05\linewidth]{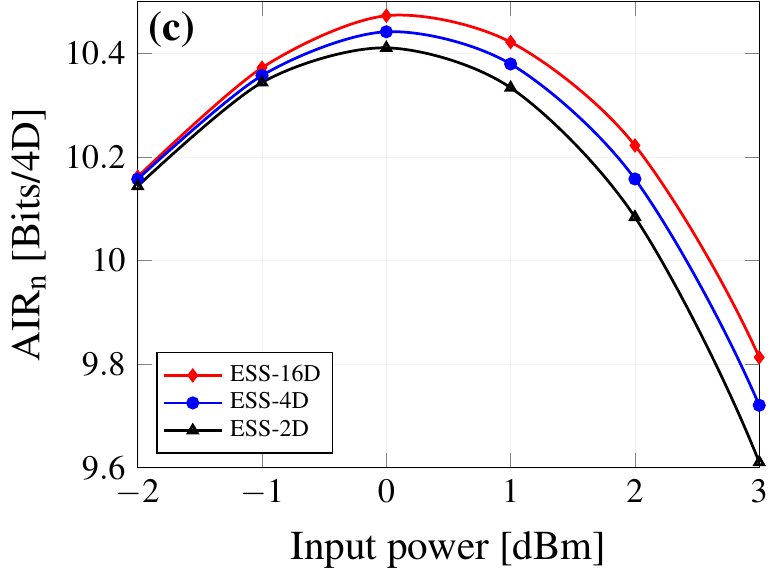}%
  \end{minipage}
  \vspace{-0.7cm}
\end{figure*}

 We numerically simulate a dual-polarization long-haul DSCM transmission system. We focus on $1$ WDM channel with $4$ digital subcarriers and a symbol rate per subcarrier of $10$ Gbaud. The subcarrier spacing is $2$ GHz and a root-raised cosine (RRC) filter, with roll-off factor of $0.1$ is applied for each subcarrier. $64$-QAM is used as modulation format for the shaping schemes and the number of transmitter symbols is $108,000$.
We consider a dispersion unmanaged link with a multi-span standard single mode fiber (SSMF) and a span length of $L=80$~km. 
The erbium-doped fiber amplifier (EDFA) noise figure is $4.5$ dB and its gain $16$ dB. The total transmission distance is $2000$ km. The receiver DSP per subcarrier consists of chromatic dispersion (CD) compensation, an RRC matched filter, and downsampling to $1$ sample/symbol before performing an ideal compensation of the common phase rotation of the entire constellation due to nonlinearity. To focus on the impact of the nonlinear effects, polarization mode dispersion and linear phase noise are not considered in this work. The performance is evaluated in terms of finite-length AIR (AIR$_n$) and GSNR.
The finite-length AIR takes into account the rate loss introduced by short shaping blocklengths. It is defined as in \cite{ps2}. The GSNR accounts for both, amplified spontaneous emission noise and nonlinear interference. It is known also as effective SNR and given as in \cite{me1}. To recover the transmitted information bits, intra-DM demapping has to be performed first to obtain the shaped amplitudes of the $4$ subcarriers. Then, deshaping can be done to recover the transmitted bits from the shaped amplitudes.

Since the proposed shaping scheme uses two polarizations and $4$ subcarriers, it generates a $16$D signal. We focus on the nonlinear behavior of the different intra-DM shaping schemes for a middle subcarrier, which is more affected by nonlinear fiber impairments compared to outer subcarriers.  
In Fig. \ref{fig:2}(b), we plot the GSNR versus input power. The proposed ESS-based $16$D scheme improves the GSNR performance at optimal input power by about $0.1$ dB and $0.21$ dB in comparison with the conventional $2$D intra-DM and the $4$D super symbol intra-DM mapping, respectively. The gain increases with input power due to the increase of the nonlinear effects. The constellation diagrams at $1$ dBm input power of the different intra-DM pairing schemes are inserted in Fig. \ref{fig:2}(b). In Fig. \ref{fig:2}(c), the finite length AIR is plotted as a function of the input power. At optimal input power of $0$ dBm, the proposed $16$D pairing provides a gain of about $0.04$ bits/$4$D and $0.07$ bits/$4$D  with respect to $2$D and $4$D super symbol pairing, respectively. The three pairing schemes have the same blocklength $n=360$, which is translated to the same shaping gain. Consequently, the performance improvement is purely due to the nonlinear tolerance gain. 


The nonlinear tolerance gain that $16$D pairing provides, can be explained by the partial mitigation of inter-subcarrier nonlinear interferences. This is due to the correlation between the subcarriers, obtained by applying the $16$D pairing. To exploit the nonlinear tolerance gain provided by $16$D pairing, pre-interleaving is required to keep the temporal structure of the signal, which is required in general for any intra-DM pairing scheme \cite{psnl1,psnl2}.   
\vspace{-0.2cm}
\section{Conclusion}
We proposed a scalable multi-dimensional symbol mapping technique using an intra-distribution matching shaper based on polarization and frequency dimensions and compared it to conventional $2$D and $4$D super symbol intra-DM shaping. We demonstrated that our proposed scheme improves the GSNR and AIR performance by partially mitigating the inter-subcarrier nonlinear interference in digital subcarrier multiplexing transmission systems.  
\vspace{-0.2cm}
\small
\flushleft
\textbf{Acknowledgment:} This work is funded by the German Ministry of Education and Research (OptiCON, grant 16KIS0993).

\end{document}